\documentclass[12pt]{article}

\usepackage{physics}
\usepackage{bigints}
\usepackage{amsthm}
\usepackage{multirow}
\usepackage{graphicx,capt-of}
\usepackage{color}
\usepackage{amssymb,amsmath,amsfonts,bm}
\usepackage{pdflscape}
\usepackage{placeins}
\usepackage{subfigure}
\usepackage{caption}
\usepackage{tabularx,ragged2e,booktabs,caption}
\usepackage{grffile}
\makeatletter
\newcommand*{\rom}[1]{\expandafter\@slowromancap\romannumeral #1@}
\makeatother

\def\beq{\begin{equation}}
	\def\eeq{\end{equation}}
\def\ba{\begin{eqnarray}}
	\def\ea{\end{eqnarray}}

\begin{document}
	
	\begin{center}
		{\Large{\bf A possible phase and dynamical transition in a three-dimensional Electron Glass}} \\
		\ \\
		\ \\
		by \\
		Preeti Bhandari$^{1}$ and Vikas Malik$^2$  \\
		$^1$Department of Physics, Ben Gurion University of the Negev, Beer Sheva 84105, Israel. \\
		$^2$Department of Physics and Material Science, Jaypee Institute of Information Technology, Uttar Pradesh 201309, India. \\
	\end{center}
	
	\begin{abstract}
Using mean-field approximations, this paper identifies a phase transition in a three-dimensional Electron Glass lattice model. The density of states of the eigenvalue distribution of the inverse susceptibility matrix  is used to identify the possibility of a phase transition. In the thermodynamic limit, the eigenvalue spectrum appears to extend to zero as $T \approx T_{c}$. To determine the dynamical relaxation laws near the transition temperature, we use the eigenvalue distribution of the linear dynamical matrix. Our analysis distinguishes between the phenomenon of phase transition and slow dynamics.
	\end{abstract}
	
	\newpage
	
	\section{Introduction}
	\label{intro}
	
A system of localized electrons interacting via a long-range Coulomb potential is called an electron glass (EG). It is commonly realized at low temperatures in impurity-doped semiconductors when the Fermi level is in the impurity band. The exceedingly interesting interplay between strong interactions and disorder that EG exhibits presents a formidable theoretical challenge. Particularly significant is the fact that the long-range Coulomb interaction is not screened, resulting in the so-called Coulomb gap in the density of single-particle states (DOS) close to the Fermi level \cite{efroscoulomb1975}. This indicates that the low-temperature DOS diminishes toward Fermi-level ($\mu$) but only completely disappears at $T = 0$ and $E = \mu$. Efros and Shklovskii \cite{efroscoulomb1975} developed a self-consistent equation (SCE) and predicted that the zero-temperature single-particle DOS $g(E, T=0)$ in the Coulomb gap follows a universal power law $g(E,0) \sim |E - \mu|^{D-1}$ (where $D$ is the spatial dimension) based on the stability criteria of the ground state against one-particle hops. The analytical theories \cite{mogilyanskiiself1989,vojtabethe1993} state that the finite-temperature DOS at the Fermi level is proportional to zero-temperature DOS at an energy $k_{B}T$ distant from the Fermi energy i.e., $g(\mu,T) \sim g(E,T=0)$ with $|E - \mu| = k_{B}T$. Consequently, $g(\mu,T) \sim T^{D-1}$ is predicted by the BPW approximation \cite{vojtabethe1993} as well as the SCE technique \cite{efroscoulomb1975,efroscoulomb1976}. 

Over the past few decades, there has been substantial research into the thermodynamics of long-range spin glasses (SG), which are analogous to the electron glass model. The analysis of numerous time-dependent susceptibilities \cite{cannellamagnetic1972,muldersusceptibility1981,tholencefrequency1980}, such as the dependence of remanent magnetization on magnetic history, demonstrates the close connection between the static and dynamic characteristics of spin glasses \cite{binderspin1986,hertzspectra1983,hertzdynamics1979,kinzelstatic1983,kumarrelaxation1980,sompolinskyrelaxational1982}. Thus, dynamic processes must be understood to comprehend the equilibrium characteristics of a glassy system. 

Even though the study of the glass transition has made great strides, \cite{davieselectron1982,grunewaldmean1982,pollakcoulomb1982,pollaknon1984,grannangrannan1994} it still is an open question. Some mean-field investigations \cite{muellerglass2004,pastormelting1999,pankovnonlinear2005,muellermean2007,brayspin1982} suggest the occurrence of a stable glassy phase (also supported by recent numerical analysis \cite{barzegarnumerical2019}), whereas others contradict the possibility of a glass transition in three-dimensional EG \cite{goethephase2009,surerdensity2009,mobiuscomment2010}. Consequently, there is an ongoing investigation into the possibility of a glass transition.

In this paper, we show there is a phase transition in the three-dimensional lattice Electron Glass model under mean-field approximation. In the thermodynamic limit at $T = T_{c}$, we see a phase transition, indicated by the eigenvalue spectrum of the inverse susceptibility matrix appearing to extend down to zero. We also find that the eigenvector corresponding to the smallest eigenvalue at $T_{c}$ is extended. To investigate the dynamics, we use the mean-field approximation to solve the master equation and compute the linear-dynamical matrix (``A-matrix"). The relaxation dynamics of the system at finite temperature is described using the eigenvalue distribution of the ``A-matrix".

The outline of the paper is as follows: Sec.(\ref{sec2}) provides a detailed discussion about the dynamics of the Electron Glass model starting from the master equation. In Sec.(\ref{sec3}), the results are discussed, starting with the brief discussion on Coulomb gap results in Sec.(\ref{sec3a}). In Sec.(\ref{sec3b}), the probability distribution of eigenvalues of the inverse susceptibility matrix is investigated, and finally Sec.(\ref{sec3c}) outlines how the relaxation dynamics of the system affects the phase transition. Sec.(\ref{sec4}) provides a summary of the findings and proposes suggestions for further investigations. 
\section{Theoretical background}\label{sec2}
\subsection{Model}
\label{theo_model}
We investigate a lattice of $N-$sites, and each site has a localized energy level with random energy $\phi_{i}$. The electrons interact via unscreened Coulomb interactions, $K_{ij} = \frac{e^{2}}{r_{ij}}$. The Hamiltonian for the EG model can be written as

\begin{equation}
	\label{Hamil}
	\mathcal{H} = \sum_{i = 1}^{N} \phi_{i} n_{i} + \frac{1}{2} \sum_{i \neq j}^{N} K_{ij} \Big(n_{i} - \frac{1}{2}\Big) \Big(n_{j} - \frac{1}{2}\Big)
\end{equation}
where $\phi_{i}$ is random energy with probability distribution $P(\phi_{i})$. $P(\phi_{i})$ is typically a box or Gaussian distribution. The important parameter for theory is the width of the disorder distribution ($W$), which signifies the strength of disorder. The occupation number $n_{i} \in \{0,1\}$.

\subsection{Kinetics of Electron Glass}

The Hamiltonian described in Eq.(\ref{Hamil}) has no intrinsic dynamics. Physically, electrons hop from one localized state to another by the mediation of phonons, which also provide the energy required in the transition (or take up energy released). Using Fermi golden rule, Miller and Abraham \cite{millerimpurity1960} derived the phonon-assisted transition probability from a site $i$ to site $j$ as

\begin{subequations}
	\label{rate_eq1}
	\begin{equation}
		W_{ij} = \nu_{0}\, exp \Bigg( \frac{-2r_{ij}}{\xi}\Bigg) N_{B}(\Delta_{ij}) \quad \quad  when \, \Delta_{ij} > 0 
	\end{equation}
	\begin{equation}
		W_{ij} = \nu_{0}\, exp \Bigg( \frac{-2r_{ij}}{\xi}\Bigg) [1 + N_{B}(\Delta_{ij})] \quad \quad  when \, \Delta_{ij} < 0 		
	\end{equation}
\end{subequations}
where $\xi$ is the localization length, $r_{ij}$ is the distance between sites $i$ and $j$ and $\nu_{0}$ is the phonon frequency and $N_{B}(\Delta_{ij}) = [exp(\Delta_{ij}/k_{B}T)^{-1}]$ is the probability of absorbing the phonon of energy $\Delta_{ij}$ which is proportional to the number of phonons available at energy $\Delta_{ij}$. The first exponential factor in equations (\ref{rate_eq1}(a)) and (\ref{rate_eq1}(b)) is due to the tunneling of an electron between sites $i$ and $j$ and is just the overlap between wave functions at the two sites.

The dynamics governed by the transition rates defined in Eq.(\ref{rate_eq1}) is stochastic and is described by the master equation in the probability space. We introduce the probability distribution $P(n_{1}, n_{2},...,n_{N},t)$, which is the probability for the $N-sites$ taking some set of values of $\{n_{i}\}$ at time $t$. It is the generalization of the distribution for a single particle, for which we have two probabilities, $P(1,t)$ and $P(0,t)$.

Next, we model how this probability changes over time. This can be done in a number of ways, and the choice among these is made on the basis of the physical processes involved. The simplest way was introduced by Glauber \cite{glaubertime1963}, who assumed that dynamics proceeds by single spin flips. But, the dynamics of electron hops is not described by Glauber dynamics as this dynamics changes the occupation of just one site at a time and hence does not conserve the number of electrons. The Kawasaki formulation \cite{kawasakikinetics1972} is applicable to the electron glass as the electron number is conserved, which is equivalent to fixed magnetization. 

In the EG model, a site can be occupied by only one electron. The occupation probability $f_{i}$ of a site $i$ is determined by the rate equation

\begin{equation}
	\label{rate_eq2}
	\frac{df_{i}}{dt} = \sum_{j} [f_{j} (1 - f_{i}) W_{ji} - f_{i} (1 - f_{j}) W_{ij} ]
\end{equation}
where $W_{ij}$ is the one-electron transition rate from a site $i$ to $j$ as defined in Eq.(\ref{rate_eq1}). In thermal equilibrium, the rate equation should obey the detailed balance condition, and $f_{m}$ takes the Fermi-Dirac form

\begin{equation}
	\label{FD_eq}
	f_{m}^{0} = \frac{1}{exp[\beta(E_{m} - \mu)] + 1}
\end{equation}
where $E_{m}$ is the energy of an electron on the site $m$, $\mu$ is the Fermi energy and $\beta = (k_{B}T)^{-1}$. The energy of the electron at site $i$ in equilibrium is defined as

\begin{equation}
	\label{HE_eq}
	E_{i}^{0} = \phi_{i} + \sum_{j \neq i} K_{ij} f_{j}^{0} \, .
\end{equation}
Equations (\ref{FD_eq}) and (\ref{HE_eq}) are solved self consistently to obtain the equilibrium occupation probabilities ($f_{i}^{0}$).

A lot of insight can be gained by studying a linearized version of the rate equations. By linearizing $f_{i}(t)$ about the equilibrium values of $f_{i}^{0}$, we get
\begin{equation}
	\label{eq2}
	f_{i}(t) = f_{i}^{0} + \delta f_{i}
\end{equation}
And the final linear equation using the detailed balance is:
\begin{equation}
	\label{eq18}
	\dfrac{d}{dt} \delta f_{i}= \sum_{l} A_{i l} \hspace*{2mm} \delta f_{l}
\end{equation} 
where 
\begin{subequations}
	\label{19}
	\begin{equation}
		A_{i i}  = -\sum_{k \neq i}\, \frac{\Gamma_{i k}}{f_{i}(1-f_{i})} 
	\end{equation}
	\begin{equation}
		A_{i l} = \frac{\Gamma_{l i}}{f_{l}(1-f_{l})} + \frac{1}{T} \sum_{k(\neq l\neq i)}\, \Gamma_{i k} \hspace*{2mm} (K_{k l}-K_{i l})
	\end{equation}
\end{subequations}
Here we define the transition rates ($\Gamma_{i k}$) as
\begin{subequations}
	\label{Tij}
	\begin{equation}
		\Gamma_{ik} = \frac{1}{2 \tau} \gamma(r_{i k}) \hspace*{2mm} f_{i}^{0} (1-f_{k}^{0}) \hspace*{2mm} f_{FD}(E^{0}_{k} - E^{0}_{i})
	\end{equation}  
	\begin{equation}
		\Gamma_{ k i} = \frac{1}{2 \tau} \gamma(r_{k i}) \hspace*{2mm} f_{k}^{0} (1-f_{i}^{0}) \hspace*{2mm} f_{FD}(E^{0}_{i} - E^{0}_{k})
	\end{equation}
\end{subequations} 
Here $f_{FD}$, is the Fermi-Dirac distribution as defined in Eq.(\ref{FD_eq}). Thus Eq.(\ref*{19}) gives us the linear dynamical matrix ($A-matrix$). For relaxation to equilibrium, all its eigenvalues are positive. 
In the particle language, we define susceptibility as $\chi_{im} = - \partial f_{i}/ \partial \phi_{m} $, so the inverse susceptibility matrix becomes

\begin{equation}
	\label{eq4}
	\chi^{-1}_{il} = \frac{\delta_{il}}{\beta f_{i} (1-f_{i})}  +  K_{il}
\end{equation}

The ``A-matrix" is related to the susceptibility matrix by the relation 
\begin{equation}
	\label{eq41}
	A_{ij} = - \beta \sum \gamma_{il} \,  \chi^{-1}_{lj} 
\end{equation}
where
\begin{equation}
	\label{eq42}
	\gamma_{il} = - \Gamma_{il} + \delta_{il} \sum_{k} \Gamma_{i k}
\end{equation}
The important point for us is how the eigenvalues of the ``A-matrix" and the $``\chi^{-1}-matrix"$ behave as $T \leq T_{c}$. 
\section{Results}
\label{sec3}
\subsection{Coulomb gap}
\label{sec3a}
The computation details are as follows: The calculations are carried out on a cubic lattice of $N = L \times L \times L$ sites with periodic boundary conditions. A random energy $\phi_{i}$ is allocated to each site in accordance with the box-distribution $[-W/2,W/2]$ where $W = 6$. In the case of half-filling, initially $N/2$ electrons are randomly distributed across the $N$ sites. To obtain equilibrium magnetization under mean-field approximations the following equation was solved self-consistently,

\begin{equation}
	\label{mag_mft}
	m_{i} = tanh \ \beta \ \Bigg(\varepsilon_{i} + \sum_{k} \frac{m_{k}}{r_{ik}}\Bigg)
\end{equation}
Here $r_{ik}$ is the distance between sites $i$ and $k$. A single iteration consists of application of Eq.(\ref{mag_mft}) with $i = 1,2,. . . , N$ successively. Convergence was taken to have occurred when $\sum_{i} (m_{i}^{n+1} - m_{i}^{n})^{2}/((\sum_{i} m_{i}^{n})^{2}) < 10^{-6}$. The above equation was solved self-consistently for the system sizes  $L = 16, 12$ and $8$ and the final $m_{i}$'s were then used to calculate the Hartree energy $E_{i} = \varepsilon_{i} + \sum_{i \neq j} f_{j}/r_{ij}$ (where $f_{j} = (m_{j}+1)/2$). We have annealed our data from $\beta = 1$ to $\beta = 10$ ($\beta = 1/T$ is the inverse of temperature). Results were obtained after averaging over 100, 500 and 1000 realizations, respectively for $L = 16, 12$ and $8$.

\begin{figure}
	\centering
	\includegraphics[scale=0.7]{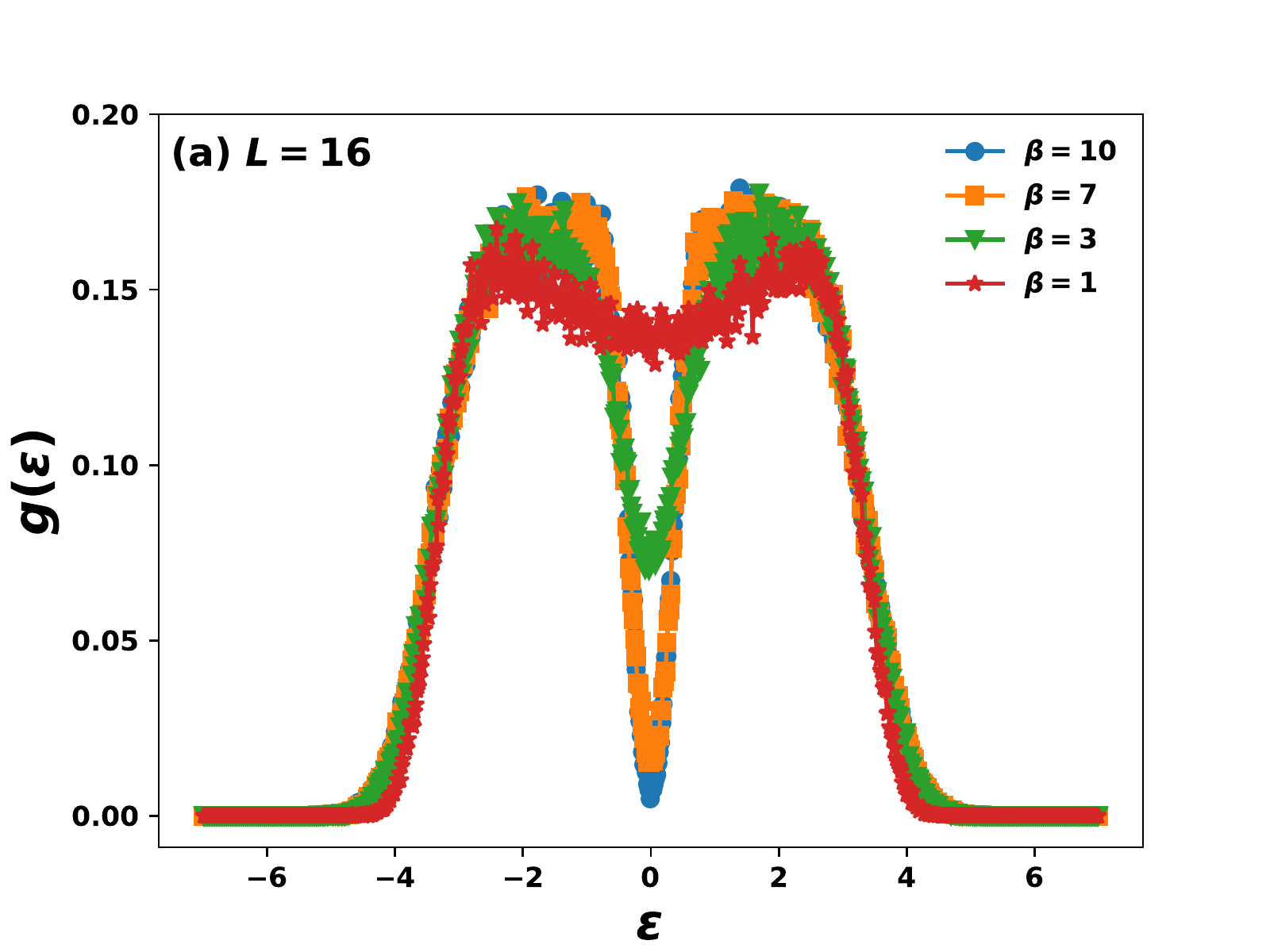}
	\includegraphics[scale=0.7]{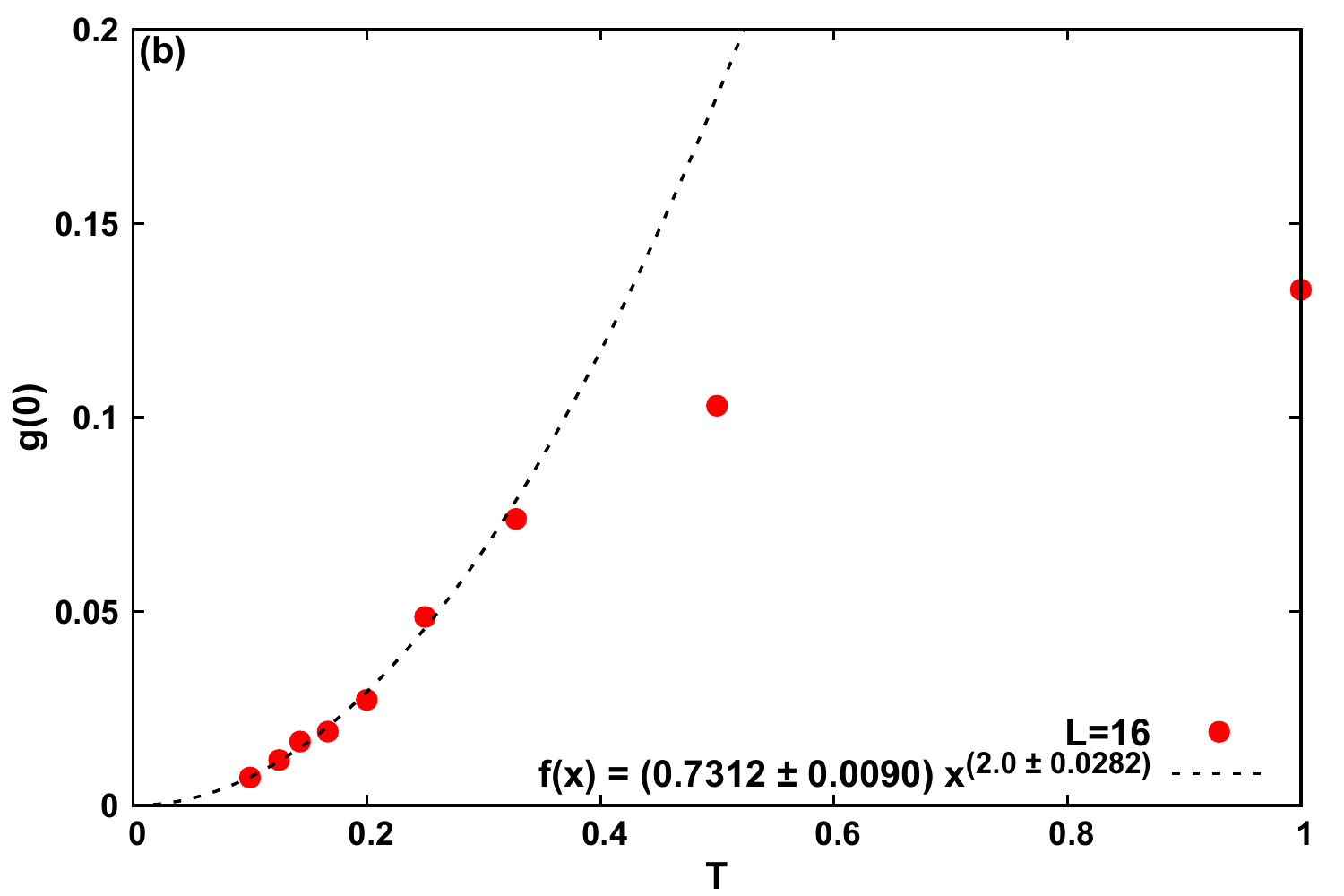}
	\caption{\label{dos_T} (a) Single particle density of states (DOS) at different temperatures for disorder value $W = 6$. (b) For the system size $N = L^{3}$ sites, DOS at the chemical potential $g(0)$ as a function of temperature (T) is plotted. The data is approximated by the functional form $g(0) = a T^2$ where $a = 0.7312 \pm 0.0090$ as shown by the solid line.  }
\end{figure}

As was previously mentioned, a soft gap, often referred to as the Coulomb gap, can be seen in the single-particle density of states at low temperatures \cite{efroscoulomb1975}. In Fig.\ref{dos_T}(a), we plot the single-particle density of states $g(E)$ vs $E$ at different temperatures, and the filling up of the Coulomb gap can be seen as the temperature increases. In Fig.\ref{dos_T}(b), we have shown a variation of $g(\mu,T)$ with temperature. When we approximated the data using $T^{2}$ functional form, we saw a little deviation from $g(\mu,T) \propto T^{d-1}$ law ($d = 3$ here) for temperatures where the gap is well formed. The temperatures where the gap is noticeably filled show a marked deviation from this law. 	

 \subsection{Inverse susceptibility matrix}
\label{sec3b}
\begin{figure}
	\centering
	\includegraphics[scale=0.5]{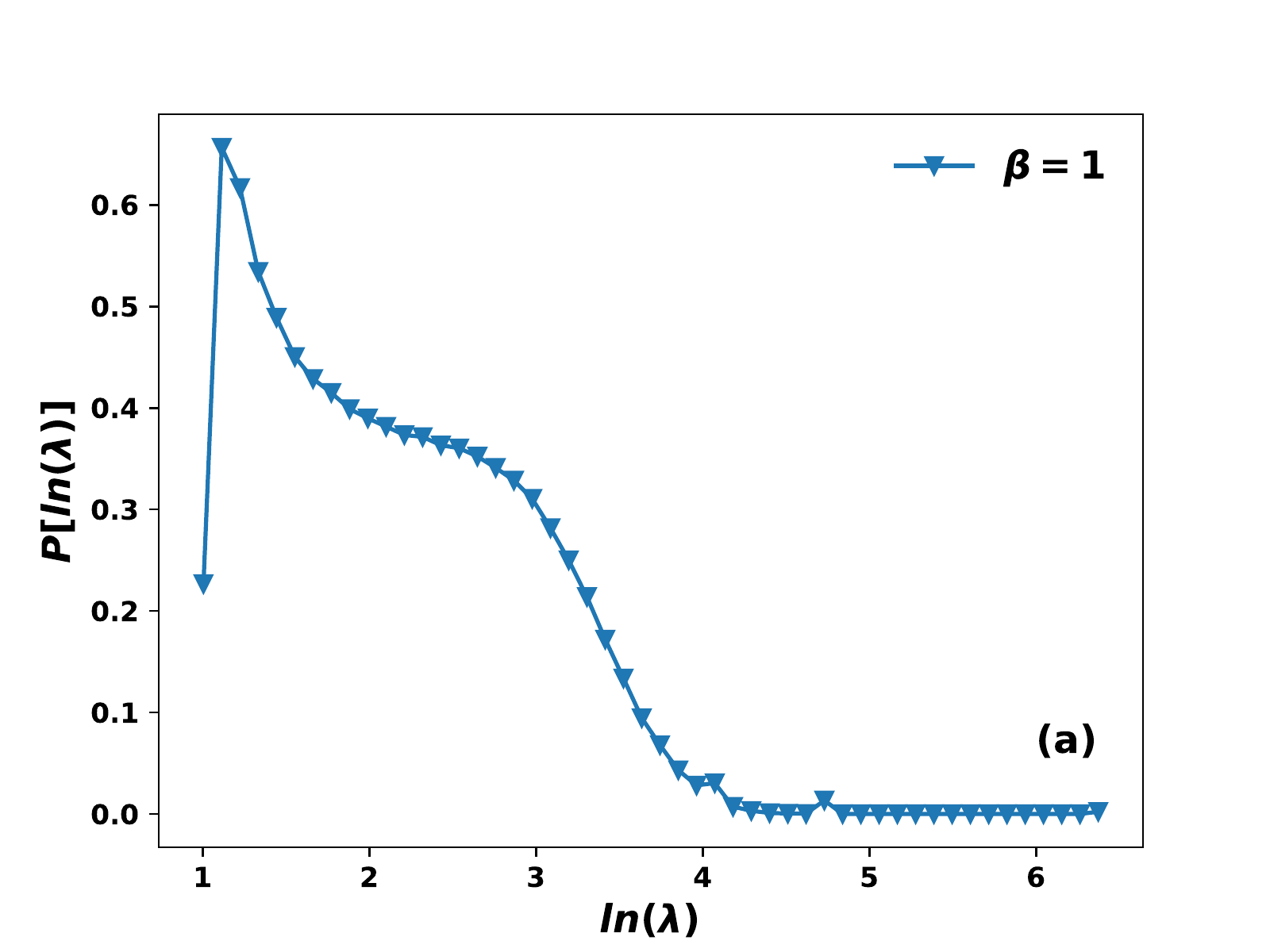}
	\includegraphics[scale=0.5]{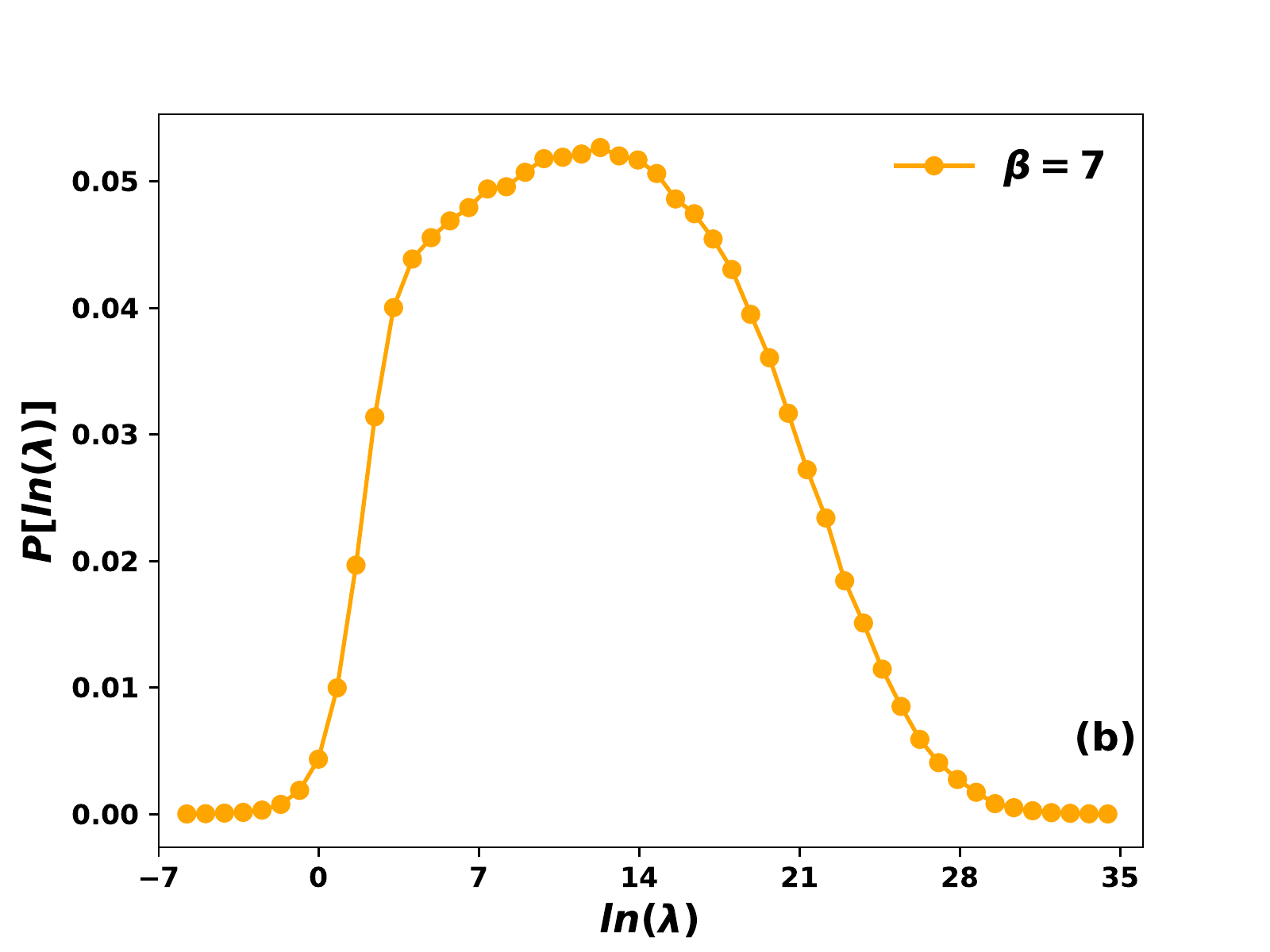}
	\includegraphics[scale=0.5]{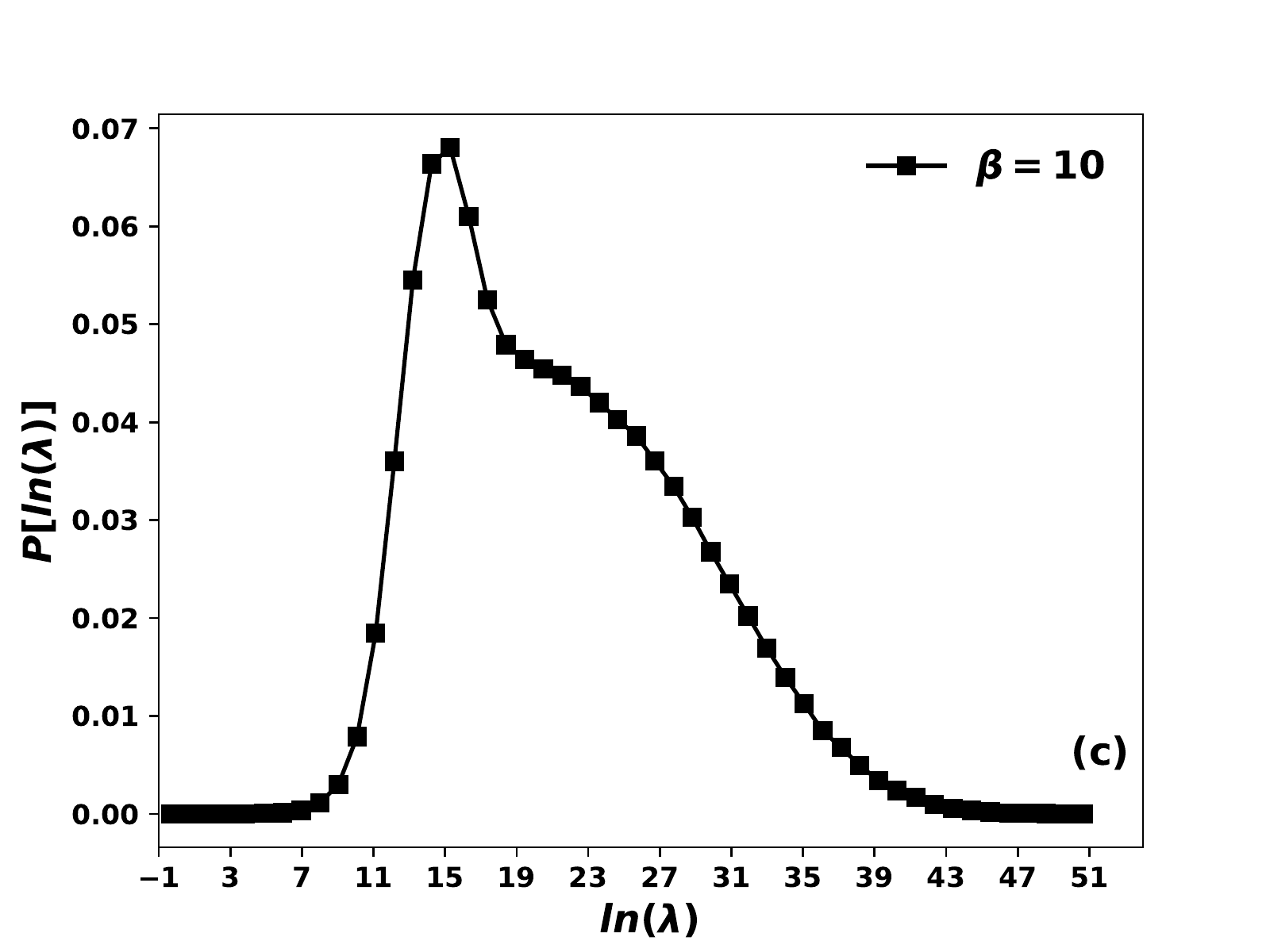}
	\caption{\label{lambda_dist} (a)-(c) Histogram of the density of eigenvalues of the inverse susceptibility matrix $\chi^{-1}$,  $P(\lambda)$, versus $\lambda$ on a log-log scale for a typical system with $N = 4096$ at different temperatures. }
\end{figure}

The eigenvalue spectrum of the inverse susceptibility matrix $\chi^{-1}$, at different temperatures, is shown in Fig.(\ref{lambda_dist}). The figure shows density of states of the eigenvalue of the inverse susceptibility matrix. At high temperatures ($\beta = 1$), eigenvalues are all well above zero, but as the temperature is lowered, the whole spectrum moves towards zero. As the temperature is lowered further ($T < T_{c}$), the eigenvalue spectrum again starts moving away from zero. For a pure system, when the smallest eigenvalue becomes zero and the system undergoes a phase transition in the mode corresponding to the smallest eigenvalue. As this mode condenses (i.e. order sets in), all the eigenvalues are renormalized by the condense. 

In contrast, for a disordered system, one can encounter a different scenario. The inverse susceptibility matrix is disordered, so its eigenfunctions can be extended or localized. The typical spectrum has mobility edges so that very large or very small eigenvalue states are localized at $T \gg T_{c}$. When the temperature is lowered, it can well happen that the eigenfunction with zero eigenvalue is localized. Condensation in this mode can not lead to a phase transition as it affects only a small part of the system. Thus the smallest eigenvalue must get renormalised and affect all other through a nonlinear effect. The transition will occur only when an extended eigenstate reaches zero. Some interesting results, which are rigorous, were derived by Bray and Moore \cite{brayevidence1979} where they argued that if the smallest eigenvalue vanished at some finite temperature, the corresponding eigenvector could not be localized.

\begin{figure}[t]
	\centering
	\includegraphics[scale=0.50]{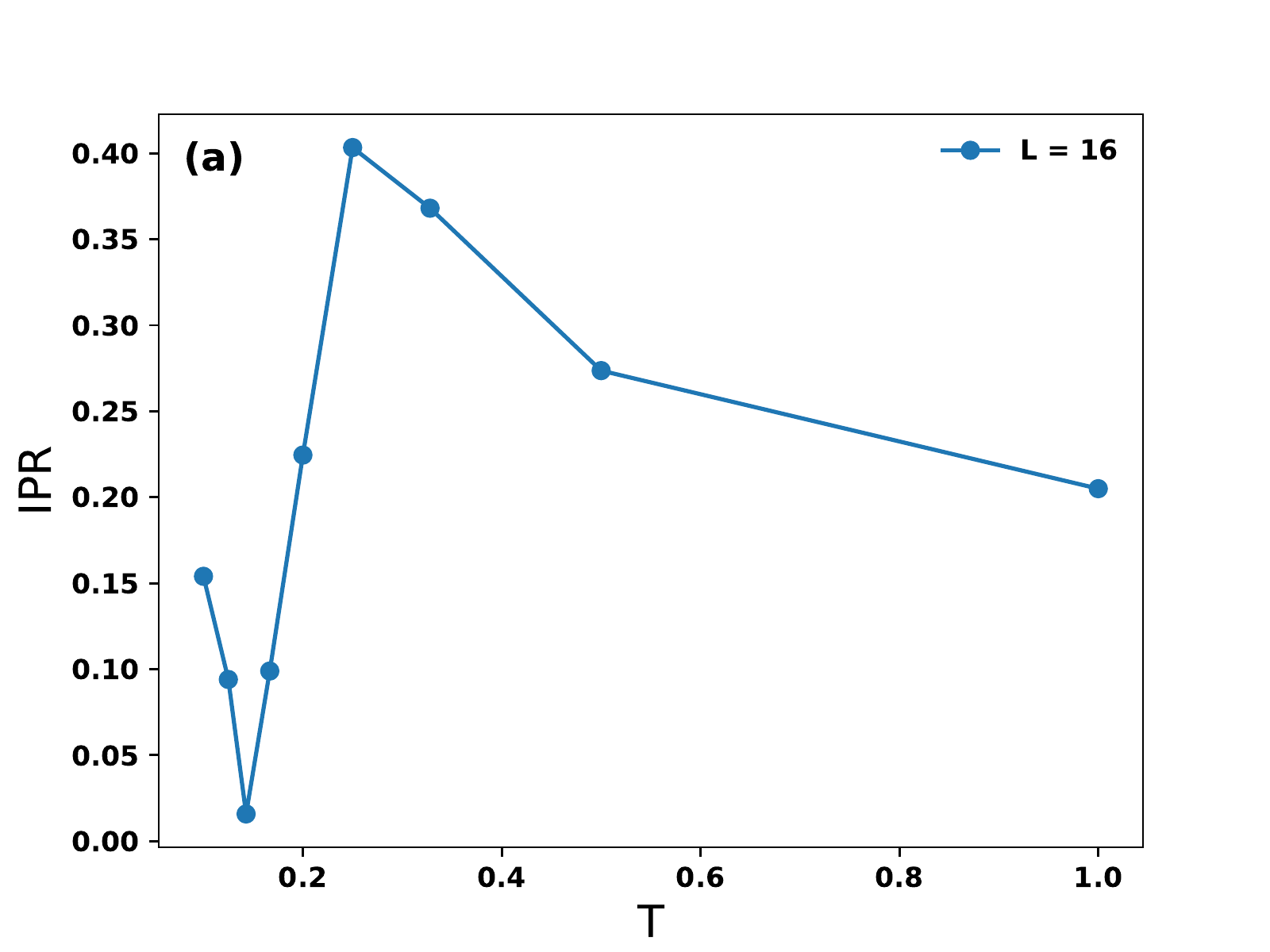}
	\includegraphics[scale=0.50]{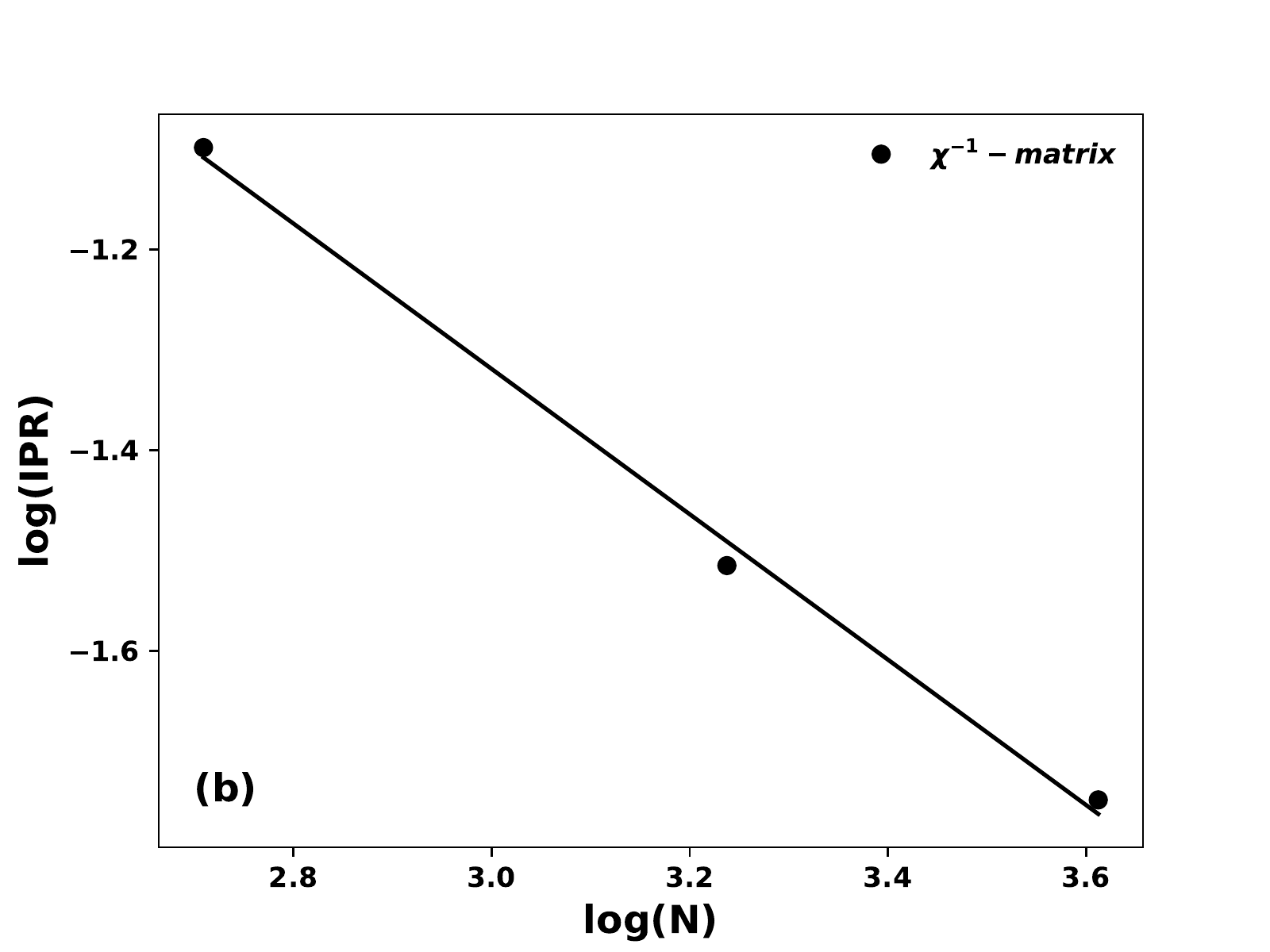}
	\includegraphics[scale=0.50]{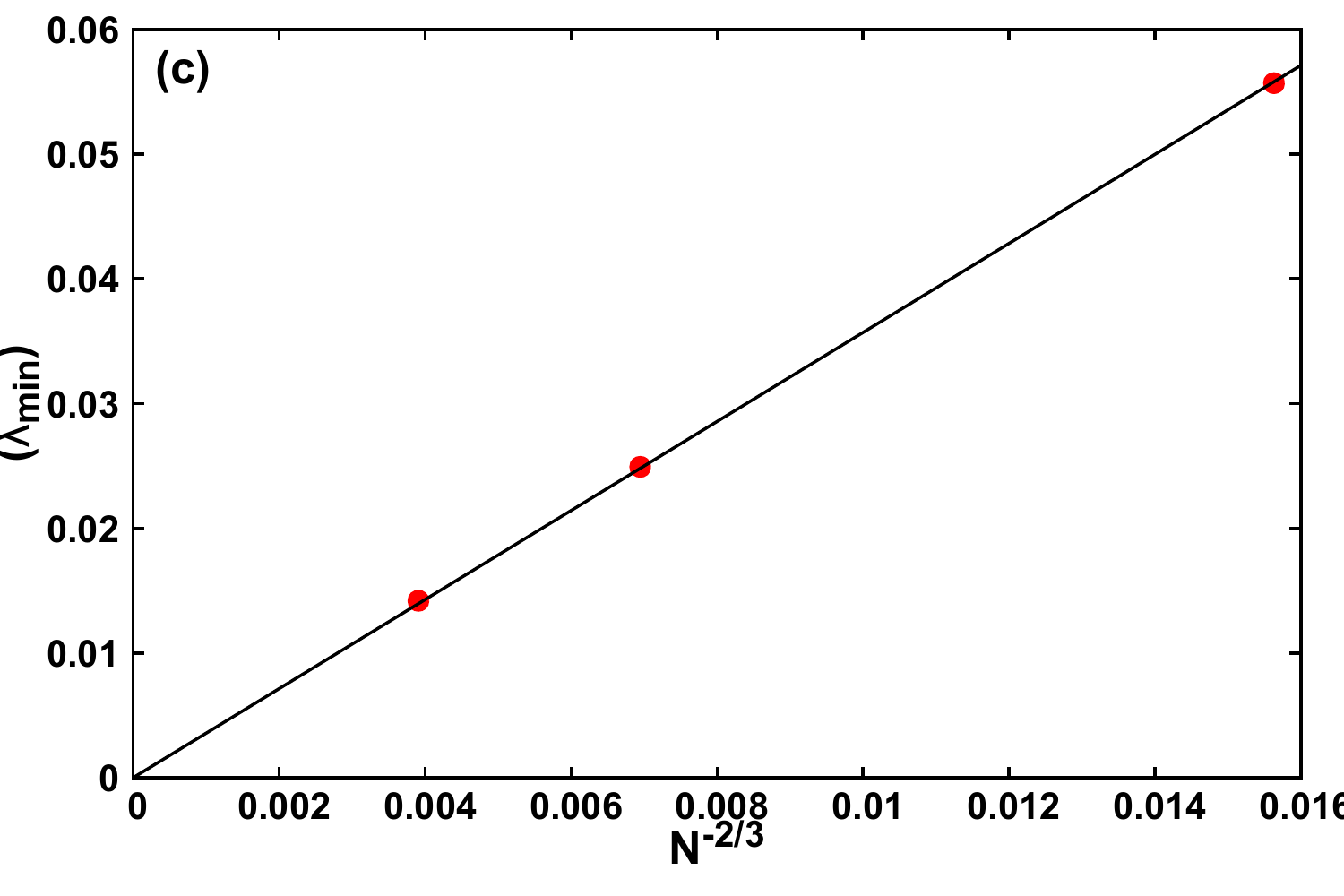}
	\caption{\label{bray_val}(a) Inverse participation ratio as a function of temperatures for system size $N = 4096$. (b) Log-log plot of the inverse participation ratio (IPR) of the inverse susceptibility matrix and the system size $N$. The solid line represents the linear fit with slope $-1$. (c) Minimum eigenvalue of the inverse susceptibility matrix vs $N^{-2/3}$.   }
\end{figure}

We also investigated the degree of localization of the eigenstate of the minimum eigenvalue ($\lambda_{min}$) of $X^{-1}-matrix$ at each temperature for various system sizes. This is accomplished by computing the Inverse Participation Ratio (IPR), which assesses the degree of localization in the system,

\begin{equation}
	\label{eq5}
	IPR (\lambda) = \sum_{i}  | \bra{i}\ket{\lambda}| ^{4}  
\end{equation}
where $\ket{\lambda}$ is the eigenvector of the inverse susceptibility matrix. The critical temperature ($T_{c}$) for system size (N) is provided by the minimal value of IPR vs T. From the data for $N = 4096$ sites shown in Fig.\ref{bray_val}(a)), we found $\beta_{c} = 7$. In addition, we demonstrate how IPR behaves in relation to system size $N$ in Fig.(\ref{bray_val}(b)). Our data shows that $IPR \sim 1/N$ which corresponds to an extended state.

In a finite system, the minimum eigenvalue cannot become zero. Finite-size scaling arguments for the spin glass system \cite{brayevidence1979} claim that at $T = T_{c}$, the smallest eigenvalue of the $\chi^{-1}-matrix$ scales with $N$ as $N^{-2/3}$. On the assumption that this is also true for our system, we plotted $\lambda_{min}$ against $N^{-2/3}$ for $N = 4096, 1728$ and $512$, at respective $T_{c}$ for each $N$ (see Fig.(\ref{bray_val})(c)). The statistics certainly support the premise $\lambda_{min} \sim N^{-2/3}$ and this implies that $\lambda_{min} \rightarrow 0$ as $N \rightarrow \infty$.

\subsection{Linear dynamical matrix}
\label{sec3c}
\begin{figure*}[t]
	\centering
	\includegraphics[scale = 0.7]{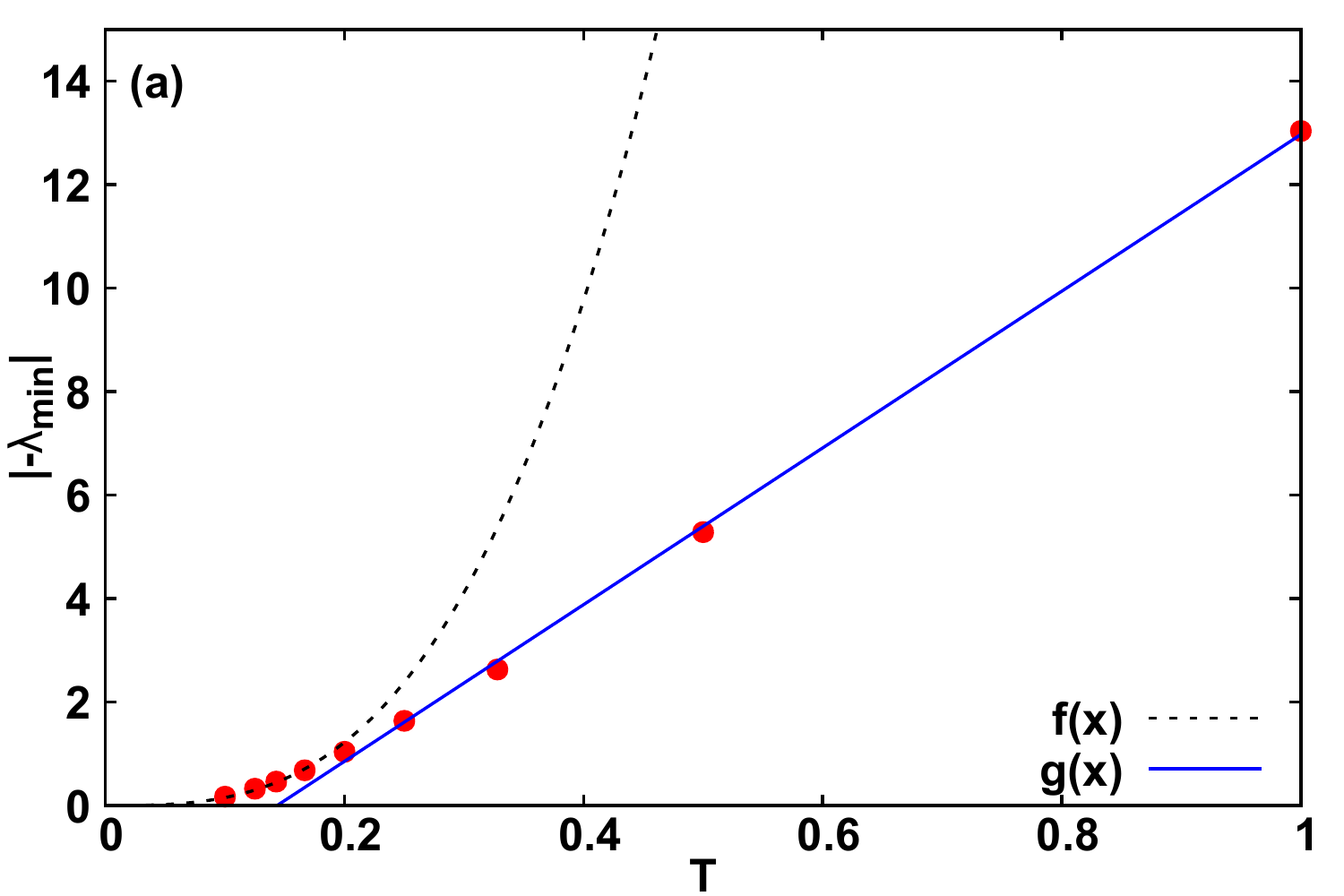}
	\includegraphics[scale = 0.7]{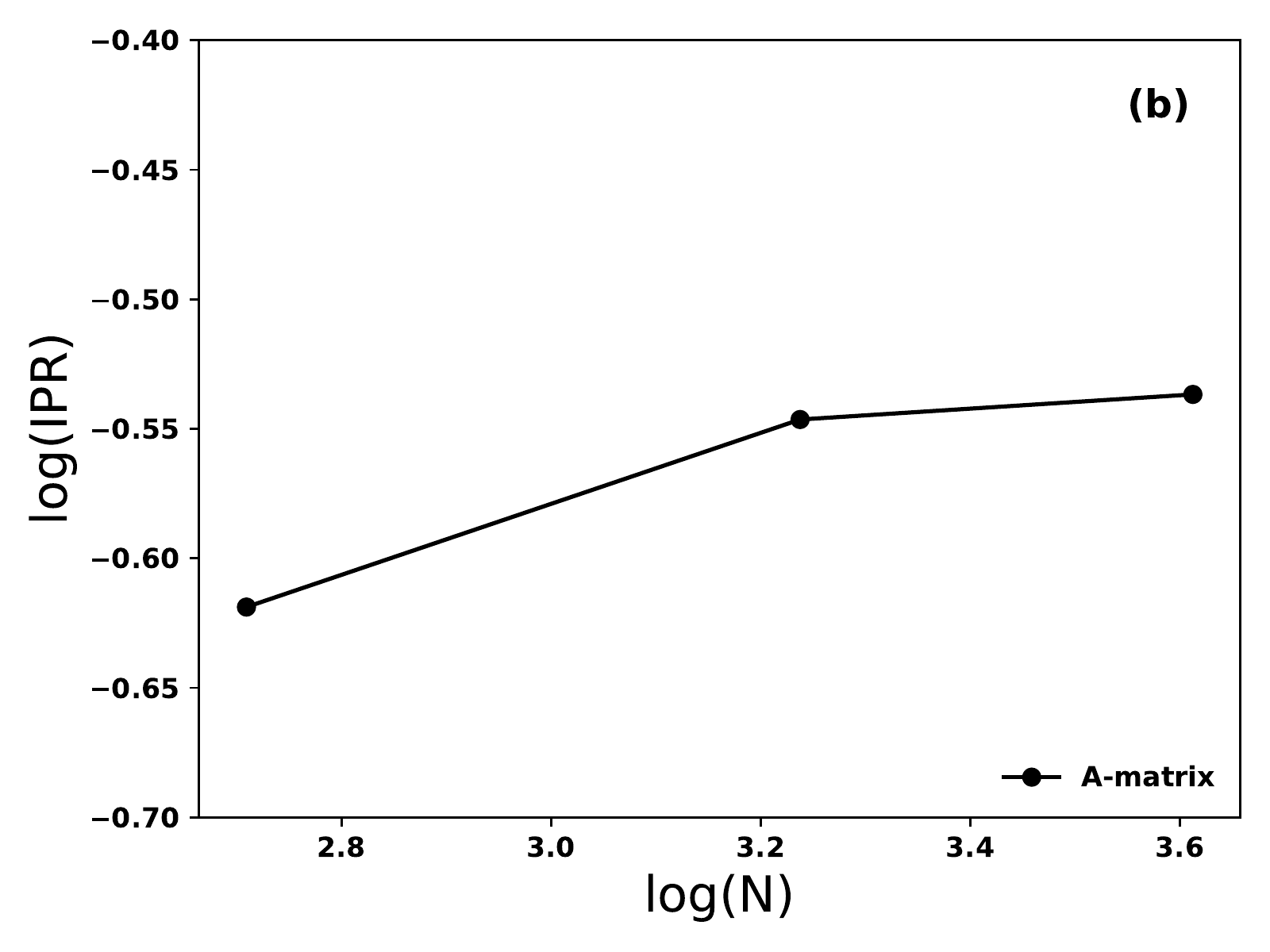}
	\caption{\label{Amat} ((a) Minimum eigenvalues of the linear dynamical matrix A at different temperatures for system size $N = 4096$, $W = 6$ and localization length $\xi = 1$ . The dashed line shows a power-law decay in the eigenvalues with temperature, $|\lambda_{min}| = aT^{3}$ where $a = 152.5470 \pm 1.3340$. The solid line shows a linear decay in the eigenvalues with temperature, $|\lambda_{min}| = aT+b$ where $a = 15.1435 \pm 0.2484$ and $b = 2.1735 \pm 0.1342$. (b) Log-log plot of the inverse participation ratio (IPR) of the linear dynamical matrix as a function of system size $N$ at $W = 6$. at $T_{c}$.}
\end{figure*}

In the previous section, we demonstrated phase transition in the EG model. One should ask what are the dynamical relaxation laws around the transition regime. We now present the numerical result for the eigenvalue distribution of the linear dynamical matrix at various temperatures. The eigenvalues here determine the rate of decay in the system. With the decrease in temperature, the shifting of the lowest eigenvalue ($\lambda_{min}$) towards zero indicates a slowing down of relaxation. It is seen that $\lambda_{min}$ follows the $T^{3}$ curve close for temperatures $T < T_{c}$. Above $T_{c}$, the minimum eigenvalue $\lambda_{min}$ is approximately linear with temperature (see Fig.\ref{Amat}(a)). This can be explained as follows:  For $T \leq T_{c}$, the Coulomb gap is well formed, and the density of states around the Fermi-level obeys $g(E) \propto E^{2}$ law (see Fig.(\ref{dos_T})(b)). The $\lambda_{min}$ corresponds to the relaxation of states around the Fermi-level. These states are localized, as can be seen from Fig.(\ref{Amat}(b)). The value of $A_{ii}$ for $E_{i} \approx 0$ (where $f_{i} = 1/2$) is given by the expression,

\begin{equation} \label{Aii_eq}
	\begin{split}
		A_{ii}^{min} & = 4 \sum_{k} \Gamma_{i k} \\
		& = 4 \sum_{k} e^{-\beta E_{k}} \gamma_{ik} \\
		& = 4 \int exp(\beta E) g(E) dE
	\end{split}
\end{equation}
which gives $A_{ii}^{min} \propto T^{3}$. For $\beta \leq 4$, the Coulomb gap fills up and  $A_{ii}^{min} \propto T$. For $t > \tau = 1/\lambda_{min}$, fluctuations ($\delta f(t)$) decay via exponential decay law, $\delta f(t) \propto exp(-t/\tau)$. One can use $A_{ii}^{min}$ to estimate $\lambda_{min}$. Thus according to our model, this crossover to exponential  decay depends upon the temperature of the system. At higher temperatures where the gap is filled $\tau \sim 1/T$ and at temperatures where the gap is well formed $\tau \sim 1/T^{3}$. In experiments, as the temperature is lowered and the gap forms, one will also see a transition from Mott's law to the ES law of conductivity.

More importantly, the localized nature of these states implies that the longtime relaxation effects are due to the sites which are not part of the percolating cluster. This implies sites that contribute to conductivity and relaxation are different.
	\section{Conclusion}
	\label{sec4}
	In this work, we present numerical evidence for a phase transition in a three-dimensional Electron Glass model. The eigenvalues of the inverse susceptibility matrix have a distribution that extends down to zero in the thermodynamic limit for $T = T_{c}$. We also confirm the fact that the lowest eigenvalue corresponds to an extended state. Our analysis of the eigenvalue distribution of the linear dynamical matrix reveals that the relaxation to the local minima is dictated by the $T^{3}$ law below the transition temperature; however, at high temperatures where the Coulomb gap is less substantial, the relaxation dynamics is faster (approximately linear with temperature.) This means for $T \ll T_{c}$, the EG system displays slow relaxation. Since this slowness depends on the density of states around the Fermi level, it will be exhibited in a two-dimensional (2D) EG model as well at a small temperature where the gap is well formed. Our analysis shows that the relaxation phenomenon is not dependent on the presence of transition, so although there is no glass transition in 2D systems, slow relaxation will occur and is in-fact observed experimentally.  
	
	\section*{CRediT authorship contribution statement}
	
	\textbf{Preeti Bhandari}: Writing original draft, Numerical computation, Analytical calculation.
	
	\textbf{Vikas Malik}: Conceptualization, Analytical calculation, Supervision, Writing, review and editing.
	
	\section*{Acknowledgments}
Vikas Malik acknowledges the funding from SERB, Department of Science and Technology, Govt of India under the research grant no: CRG/2022/004029.

	\newpage
	
	\newpage
	\bibliographystyle{unsrt}
	\bibliography{cas-refs}
	
	\newpage
	
\end{document}